\documentclass[prl,twocolumn,superscriptaddress,showpacs]{revtex4}

\usepackage{amsmath}
\usepackage{amssymb}
\usepackage{graphicx}
\usepackage{subfigure}

\newcommand{\be}[1]{\begin{equation}\label{#1}}
\newcommand{\ee}{\end{equation}}
\newcommand{\ba}[1]{\begin{eqnarray}\label{#1}}
\newcommand{\ea}{\end{eqnarray}}
\newcommand{\rf}[1]{(\ref{#1})}
\newcommand{\nn}{\nonumber}

\begin{document}

\title{Extending the range of the inductionless magnetorotational instability}

\author{Oleg N. Kirillov}
\email{o.kirillov@hzdr.de}
\affiliation{Helmholtz-Zentrum Dresden-Rossendorf\\
P.O. Box 510119, D-01314 Dresden, Germany}

\author{Frank Stefani}
\email{f.stefani@hzdr.de}
\affiliation{Helmholtz-Zentrum Dresden-Rossendorf\\
P.O. Box 510119, D-01314 Dresden, Germany}

\date{\today}

\begin{abstract}

The magnetorotational instability (MRI) can destabilize
hydrodynamically stable rotational flows, thereby allowing
angular momentum transport in accretion disks.
A notorious problem for MRI is its questionable
applicability in
regions with low magnetic Prandtl number, as they are typical
for protoplanetary disks and the outer parts of
accretion disks around black holes. Using the WKB
method, we extend the range of applicability of MRI by
showing that the inductionless versions of MRI, such as
the helical MRI and the azimuthal MRI, can easily
destabilize Keplerian  profiles $\propto r^{-3/2}$ if the radial profile of the
azimuthal magnetic field is only slightly modified from the current-free profile $\propto r^{-1}$.
This way we further show how the formerly known
lower  Liu limit of the critical Rossby number,
${\rm Ro}\approx-0.828$, connects naturally with the
upper Liu limit, ${\rm Ro}\approx+4.828$.

\end{abstract}


\maketitle

Initiated by the seminal work of Balbus and Hawley \cite{BH91},
the magnetorotational instability has become the standard
explanation for turbulence and enhanced  angular
momentum transport in accretion disks around black holes and
proto-stars. While MRI is thought to be a robust phenomenon in
the hot parts of accretion disks, a notorious problem concerns the
viability of MRI in other  regions, such as the outer parts
of black hole accretion disks \cite{BH08} and the
``dead zones'' of protoplanetary disks \cite{TURNER}.
This has to do with the fact that the onset of MRI              
demands that both  the rotation period and the Alfv{\'{e}n
crossing time
in vertical direction                                             
are shorter than the timescale for
magnetic diffusion  \cite{LIU2006}. For the case of a vertical magnetic field   
$B_z$ applied to a disk of height $H$ this means      
that both the magnetic Reynolds number                
${\rm Rm}=\mu_0 \sigma H^2 \Omega$ and the Lundquist number    
$S=\mu_0 \sigma H v_{A}$                                        
must be larger than one, and that $S \lesssim {\rm Rm}$     
($\Omega$ is the angular velocity,
$\mu_0$ is the magnetic permeability,
$\sigma$ the conductivity,
$v_A:=B_z/\sqrt{\mu_0 \rho}$
is the Alfv{\'{e}n velocity, with $\rho$ denoting the density).   
In a disk with given size, angular velocity, and magnetic field    
strength it is then often the spatially varying magnetic Prandtl
number ${\rm Pm}=\nu/\eta$,
i.e. the ratio of viscosity $\nu$ to magnetic diffusivity
$\eta:=(\mu_0 \sigma)^{-1}$, that determines the values of      
$\rm Rm$ and $S$, and hence the fate of MRI.                  

For the case without external $B_z$                               
things are even more complicated since the MRI-triggering
magnetic field, in this case dominated by the azimuthal
component $B_{\phi}$, must be produced in the disk itself,
very likely by some sort of an $\alpha-\Omega$ dynamo process
\cite{BRANDENBURG}.
This combined, loop-like action of MRI and self-excitation has
attracted much attention in the past, with many
open questions concerning issues of numerical convergence
\cite{FROMANG}, as well as the role of disk stratification
\cite{STRATIFICATION} and vertical boundary
conditions \cite{KAPILA}. Again,
the most  interesting case appears in
the limit of low ${\rm Pm}$. While Lesur and Longaretti \cite{LESUR}
have argued for a power-law decline of the turbulent                    
transport with decreasing ${\rm Pm}$, there are also                   
indications for the existence  of some critical  ${\rm Rm}$ in          
the order of 10$^3...10^4$ for the MRI-dynamo loop to work               
\cite{FLEMING-OISHI-FLOCK}.                                             

Exactly this  situation, characterized by low ${\rm Pm}$ and a
significant or even dominant $B_{\phi}$,
is the subject of intense theoretical and experimental
research initiated by Hollerbach and R\"udiger
\cite{HR95}. For the
ratio of $B_{\phi}$ to $B_{z}$ being
on the order of 1 and $B_{\phi}(r)\propto 1/r$, helical MRI (HMRI)
was shown to work also in the inductionless limit \cite{P11},
${\rm Pm}=0$, and to be governed by the Reynolds number
${\rm Re}={\rm Rm} {\rm Pm}^{-1}$ and the Hartmann number
${\rm Ha}=S {\rm Pm}^{-1/2}$,
quite in contrast to standard MRI (SMRI)
that is governed by ${\rm Rm}$ and $S$.

Somewhat disappointingly, a crucial limitation of this
surprising kind of MRI was identified by
Liu et al. \cite{LIU} who used a WKB approach to
find a minimum steepness of the rotation profile,
expressed by the
Rossby number ${\rm Ro}:= r(2\Omega)^{-1} \partial \Omega/
\partial r<{\rm Ro}_{\rm LLL}=2(1{-}\sqrt 2)\approx -0.828$.
This limit, which we call {\it lower Liu limit} (LLL)
in the following, implies that the inductionless
HMRI in the case when $B_{\phi}(r)\propto 1/r$ does not extend to the
most relevant Keplerian case, characterized by
${\rm Ro}_{\rm Kep}=-3/4$. In addition to the LLL,
the authors found also a second threshold of
the Rossby number, which we call the
{\it upper Liu limit} (ULL), at
${\rm Ro}_{\rm ULL}=2(1{+}\sqrt 2)\approx +4.828$. This
second limit, which implies a magnetic destabilization of
extremely stable flows with strongly increasing angular
frequency, has attained nearly
no attention up to present, but will play an important
role below.

The existence of the  LLL, together with a variety of further
predicted parameter dependencies, was confirmed in the PROMISE
experiment working with a low ${\rm Pm}$ liquid metal
\cite{PRE}. Present experimental work at the same device
aims at the characterization of the azimuthal MRI (AMRI),
a non-axisymmetric ``relative'' of
the axisymmetric HMRI, which is expected to dominate at
large ratios of $B_\phi$ to $B_z$ \cite{TEELUCK}.
However, AMRI as well as inductionless MRI modes with any azimuthal wavenumber
(which may be relevant at small values of $B_{\phi}/B_z$),
seem also to be  constrained by the LLL as recently shown in a
unified WKB treatment of all inductionless versions of MRI
\cite{APJ12}.
Actually, it is the apparent failure of HMRI, and AMRI,
to apply to Keplerian profiles that has prevented a wider
acceptance of those inductionless forms of MRI in the
astrophysical community.
Only recently, the intricate, though
continuous, transition between SMRI and HMRI was explained
in some detail by showing that it  involves a spectral
exceptional point at which the
inertial wave branch coalesces with the branch of the
slow magneto-Coriolis wave \cite{KS10}.

Given the fundamental importance of whether
any sort of inductionless MRI could possibly work in the low
${\rm Pm}$ regions of accretion disks, it is
quite natural to ask for how to extend the range
of its applicability beyond the LLL.
In a first attempt, the stringency of the LLL for $B_{\phi}(r)\propto 1/r$
was questioned by R\"udiger and Hollerbach \cite{rh07} who
had found an  extension of the LLL to Keplerian values
in global simulations when at least one of the
radial boundary conditions was assumed
electrically conducting.
Later, though,  by distinguishing between
convective and absolute instabilities for the travelling
waves such as HMRI, the LLL was vindicated even for such
modified electrical boundary conditions \cite{P11}.
A second attempt was made in \cite{KS11} treating
HMRI for non-zero, but low $S$.
It was found  that for $B_{\phi}(r)\propto 1/r$, the essential HMRI mode extends
from $S=0$ only to a value $S\approx0.618$, and allows for
a maximum Rossby number of ${\rm Ro}\approx-0.802$ which is indeed
slightly above the LLL, yet below the Keplerian value.
Close to this critical point, the essential HMRI is then
replaced by a helically modified SMRI.
A third possibility arises by noting that the saturation of
MRI could lead to modified flow structures with parts of steeper
shear, sandwiched with parts of shallower shear
\cite{u10}.

In this Letter, we discuss another promising way of extending
the range of applicability of the inductionless versions of
MRI to Keplerian profiles, and beyond. Rather than relying on
modified electrical boundary conditions, or on
locally \textit{steepened} $\Omega(r)$ profiles, we will evaluate
$B_\phi(r)$ profiles that are \textit{shallower} than $1/r$.
The main idea  behind that is the following:
Assume that in a low-${\rm Pm}$ region,
characterized by $S<<1$ so that standard MRI is reliably
suppressed, ${\rm Rm}$ may still be  sufficiently
large for inducing
azimuthal magnetic fields, either from a prevalent axial
field $B_z$ or  by means                                     
of a  dynamo process without any pre-given   $B_z$.           
If $B_\phi$ is produced exclusively  by an isolated axial current,
we get $B_\phi\propto 1/r$. The
other extreme case, $B_\phi \propto r$, corresponds to the
case of a homogeneous axial current density
in the fluid which is already prone to
the kink-type Tayler instability \cite{SEILMAYER}, even
at  ${\rm Re}=0$.
For real accretion disks with complicated conductivity
distributions in radial and axial direction,
quite a variety of intermediate $B_\phi(r)$ dependencies between
$\propto 1/r$  and  $\propto r$ profiles is well conceivable.
Leaving those details aside, here we focus on the generic
question which deviations of the $B_{\phi}(r)$ profile from $1/r$
could make HMRI (or AMRI) a viable mechanism for destabilizing
Keplerian rotation profiles.
By defining an appropriate {\it magnetic  Rossby number} ${\rm Rb}$
we will show that the instability extends well beyond the LLL, even reaching
${\rm Ro}=0$  when going to ${\rm Rb}=-0.5$. 
Evidently, in this extreme case
of uniform rotation the only available energy source of the instability
is the magnetic field.
Most interestingly, by tracing the instability threshold  further into the     
region of positive ${\rm Ro}$ in the ${\rm Ro}-{\rm Rb}$ plane,
we find a natural connection with the
ULL whose  meaning was a somewhat mysterious conundrum up to
present.

We set out from the equations of incompressible, viscous and
resistive magnetohydrodynamics, i.e. the
Navier-Stokes equation for the velocity field $\bf u$ and
the induction equation for the magnetic field $\bf B$,
together with the continuity equation for incompressible flows
and the divergence-free condition for the magnetic field:
\begin{eqnarray}
\frac{\partial {\bf u}}{\partial t}+{\bf u} \cdot
\nabla {\bf u}&=&\frac{{\bf B}\cdot
\nabla{\bf B}}{\mu_0 \rho} -\frac{1}{\rho}\nabla \left(p{+}\frac{{\bf B}^2}{2\mu_0}\right) {+}\nu \nabla^2 {\bf u}, \\
\frac{\partial {\bf B}}{\partial t}&=&{\bf B} \cdot \nabla {\bf u}-{\bf u}
\cdot \nabla {\bf B}
+\eta \nabla^2 {\bf B}, \\
\nabla \cdot {\bf u}&=&0,\quad  \nabla \cdot {\bf B}=0.
\end{eqnarray}
We consider a purely rotational flow exposed to a
magnetic field comprising a constant axial component  and
an azimuthal one with \textit{arbitrary} radial dependence:
\be{m4}
{\bf u}_0(r)=r\,\Omega(r)\,
{\bf e}_{\phi},
\quad {\bf B}_0(r)=B_{\phi}^0(r){\bf e}_{\phi}+B_z^0 {\bf e}_z.
\ee
To study flow and magnetic field perturbations on this background
we linearize  the equations
in the vicinity of the stationary solution by assuming
${\bf u}={\bf u}_0+{\bf u}'$, $p=p_0+p'$,
and ${\bf B}={\bf B}_0+{\bf B}'$ and leaving only
terms of first order with respect to the primed quantities.
Introducing the total  wavenumber
$|{\bf k}|^2=k^2_r+k^2_z$, and $\alpha=k_z/|{\bf k}|$,
where $k_r$ and $k_z$ are the radial and
axial wavenumbers of the perturbation,
we define the viscous,
resistive, and two Alfv\'en frequencies corresponding
to $B_z$ and $B_{\phi}$:
\begin{eqnarray}
\omega_{\nu}{=}\nu |{\bf k}|^2,~~ \omega_{\eta}{=}\eta|{\bf k}|^2, ~~
\omega_A{=}\frac{k_z {B_z^0}}{\sqrt{\rho \mu_0}},~~
\omega_{A_{\phi}}{=}\frac{B_{\phi}^0}{r \sqrt{\rho\mu_0}}.
\end{eqnarray}
Then, we define the
ratio $\beta$ of the two field components,
a re-scaled azimuthal wavenumber $n$,
the Reynolds number ${\rm Re}$, and
the Hartmann number ${\rm Ha}$ as follows:
\begin{eqnarray}
\beta=\alpha \frac{\omega_{A_{\phi}}}{\omega_A},~~  n=\frac{m}{\alpha}
,~~
{\rm Re}=\alpha\frac{\Omega}{\omega_{\nu}},~~{{\rm Ha}}=
\frac{\omega_A}{\sqrt{\omega_{\nu}\omega_{\eta}}} \; .
\end{eqnarray}
The steepness of $\Omega(r)$ will be measured
by the hydrodynamic Rossby number, and  the steepness of
$B_{\phi}(r)$ by the corresponding
magnetic Rossby number:
\begin{equation}
{\rm Ro} =\frac{r}{2 \Omega} \frac{\partial \Omega}{\partial r},\quad
{\rm Rb}=\frac{r}{2 \omega_{A_{\phi}}} \frac{\partial \omega_{A_{\phi}}}{\partial r} \; .
\end{equation}

By employing the same short-wavelength (WKB) approximation as in \cite{KGD1966,APJ12},   
but now including ${\rm Rb}$, we end up with a system of 4 coupled equations
for the perturbations of arbitrary azimuthal wavenumber, yielding
the ultimate dispersion relation $\det \left( M-\lambda I \right)=0$, with
 $\lambda$ denoting the (complex) growth rate in units of $\alpha \Omega$ and
\begin{eqnarray}
M{=}\left(
  \begin{array}{cccc}
    -i n  {-}\frac{1}{{\rm Re}} & 2\alpha   & \frac{i {\rm Ha}(1{+}n\beta)}{\sqrt{{\rm Re}\rm Rm}} & \frac{-2\alpha\beta{\rm Ha}}{\sqrt{{\rm Re}\rm Rm}} \\
    \frac{-2(1{+}{\rm Ro})}{\alpha} & -i n  {-}\frac{1}{{\rm Re}} & \frac{2\beta{\rm Ha}(1{+}{\rm Rb})}{\alpha\sqrt{{\rm Re}{\rm Rm}}} & \frac{i {\rm Ha}(1{+}n\beta)}{\sqrt{{\rm Re}{\rm Rm}}} \\
    \frac{i {\rm Ha}(1{+}n\beta)}{\sqrt{{\rm Re}\rm Rm}} & 0 & -i n  {-}\frac{1}{{\rm Rm}} & 0 \\
    \frac{-2\beta {\rm Ha}{\rm Rb}}{\alpha \sqrt{{\rm Re} \rm Rm}} & \frac{i {\rm Ha}(1{+}n\beta)}{\sqrt{{\rm Re} \rm Rm}} &
\frac{2{\rm Ro}}{\alpha} & -i n  {-}\frac{1}{{\rm Rm}}\\
  \end{array}
\right),\nn\\
\end{eqnarray}
where ${\rm Rm}={\rm Re}{\rm Pm}$ is the magnetic Reynolds number.
As a first test case, this relation can be applied
to the kink-type Tayler instability that has recently been observed
in a liquid metal experiment \cite{SEILMAYER}. In the relevant
limit with  ${\rm Pm}=0$ and ${\rm Re}=0$ we deduce from the Bilharz criterion \cite{Bilharz44} the following
 condition for marginal stability:
\begin{equation}
{\rm Rb} = \frac{(1+{\rm Ha}^2(n\beta+1)^2)^2-4{\rm Ha}^4\beta^2(n\beta+1)^2}{4{\rm Ha}^2\beta^2(1+{\rm Ha}^2(n\beta+1)^2)}  \; .
\end{equation}
For ${\rm Rb}=0$, which corresponds to $B_{\phi}\propto r$,
and taking the limit $\beta \rightarrow \infty$, we obtain
$\beta {\rm Ha}=(1-(1\pm n)^2)^{-1/2}$
which would become equal to 1 for $n=\mp1$.
Translated to the real  experiment with
$k_z \approx 2.4/r$ and a very rough estimate $k_r \approx \pi/r$, we find   
a value of ${\rm Ha}_{\rm exp}:=B_{\phi}(r) r \sqrt{\sigma/\rho \nu} \approx 34$   
which is not too far from the experimentally observed value of 22               
\cite{SEILMAYER}.

Our main focus here is, however, on the limit
${\rm Re}\rightarrow \infty$ and ${\rm Ha}\rightarrow \infty$
that is  relevant for MRI.
Assuming for the moment ${\rm Pm}=0$ (which will be slightly
relaxed later), and inserting the optimal
relation between  ${\rm Re}$ and ${\rm Ha}$,
\begin{equation}
{\rm Re}=2{\rm Rb}\sqrt{3{\rm Rb}+2}(\sqrt{1+2{\rm Rb}}+\sqrt{2{\rm Rb}})\beta^3{\rm Ha}^3 \;
\end{equation}
(obtained in the manner described in \cite{APJ12}), we find  from the Bilharz criterion \cite{Bilharz44} the  dependence of the critical Rossby
number on ${\rm Rb}$, $n$, and $\beta$:
\begin{eqnarray}
{\rm Ro}_{\rm cr}^{\pm}&=&-2 +
\frac{F\pm
\sqrt{F^2-
4 \beta^2(n\beta+1)^2}}{2\beta^2(n\beta+1)^2}F,
\label{roro}
\end{eqnarray}
where $F=(n\beta+1)^2-2\beta^2{\rm Rb}$.
Note that under the assumption ${\rm Pm}=0$ the dispersion relation possesses an exact solution,
which after being expanded into the Taylor series with respect to the interaction parameter ${\rm N}={\rm Ha}^2{\rm Re}^{-1}$ in the vicinity of ${\rm N}=0$, is
\ba{gror}
\lambda &=&-i(n\pm2\sqrt{1+{\rm Ro}})-{\rm Re}^{-1}\nn\\
&-&{\rm N}\left(F\pm\frac{\beta({\rm Ro}+2)(n\beta+1)}{\sqrt{1+{\rm Ro}}} \right)+O({\rm N}^2).
\ea
At $n=0$, ${\rm Rb}=-1$ and ${\rm Re}\rightarrow \infty$ the growth rates \rf{gror} reduce to those derived in \cite{P11}. In the limit ${\rm N}\rightarrow 0$ and ${\rm Re}\rightarrow \infty$ the
stability boundary is obtained when the real part of the term linear in ${\rm N}$ vanishes. This condition leads exactly to equation
\rf{roro}, which also confirms the correct application of the Bilharz criterion.

With the goal to find extremal values of $\rm Ro$ that are compatible with   
marginal stability, we can further  optimize $\beta$ and $n$ (or $\alpha$)     
according to
\begin{equation}
\beta_{\rm opt}=\frac{-1}{n\pm\sqrt{-2{\rm Rb}}},
~~
\alpha_{\rm opt}=\left(m{+}\frac{\omega_{A}}{\omega_{A_{\phi}}}\right)\frac{\pm1}{\sqrt{-2{\rm Rb}}}
\end{equation}
to obtain  ${\rm Ro}_{\rm opt}^{\pm}({\rm Rb}){=}-2{-}4{\rm Rb}{\pm}2 (2{\rm Rb}(2{\rm Rb}{+}1))^{1/2}$,
or
\begin{equation}
{\rm Rb}=-\frac{1}{8}\frac{({\rm Ro}+2)^2}{{\rm Ro}+1}.
\end{equation}
This relation, which is the central result of this Letter,
is illustrated in Figure 1. Let us start at the LLL, i.e.  at ${\rm Rb}=-1$,
 ${\rm Ro}_{\rm LLL}={\rm Ro}_{\rm opt}^{-}(-1)\approx-0.828$. With increasing ${\rm Rb}$,
${\rm Ro}_{\rm opt}^{-}({\rm Rb})$ also increases and reaches the Keplerian value ${\rm Ro}=-3/4$ at
${\rm Rb}=-25/32=-0.78125$. At  ${\rm Rb}=-1/2$ we arrive at solid body
rotation, i.e. ${\rm Ro}=0$. Interestingly, being connected at ${\rm Rb}=-0.5$ to the branch ${\rm Ro}_{\rm opt}^{+}({\rm Rb})$ the threshold continues even into the
positive ${\rm Ro}$ region corresponding to an outward increasing angular frequency.
Finally it meets the ULL at ${\rm Ro}_{\rm opt}^{+}(-1)\approx+4.828$ when ${\rm Rb}$ comes back  to
$-1$.
\begin{figure}[htp]
    \begin{center}
    \includegraphics[angle=0, width=0.36\textwidth]{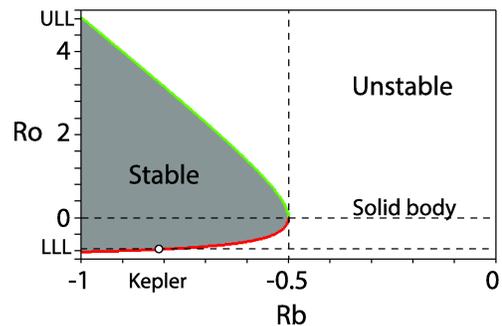}
    \end{center}
    \caption{Dependence of the optimal critical Rossby numbers ${\rm Ro}_{\rm opt}^{\pm}$ on ${\rm Rb}$ when ${\rm Pm}=0$ and ${\rm N} \rightarrow 0$.}
    \label{fig1}
    \end{figure}

Having thus seen that HMRI can easily extend to Keplerian profiles,
we still have to confirm that the shallow $B_{\phi}(r)$ profiles can indeed be
produced by induction effects for which some finite value
of ${\rm Rm}$ is still necessary. For the sake of illustration,                         
we choose now ${\rm Ro}_{\rm Kep}=-3/4$, and ${\rm Ha}=30$. Figure 2 shows two groups   
of critical curves in the $\beta-{\rm Pm}$ plane. The four curves on the right side
correspond to SMRI, the  curves continuing into the left part correspond to
HMRI. The latter ones consist, in general, of two parts, one reaching the
inductionless ${\rm Pm}=0$ area. The connection between them typically happens at
${\rm Rm}\propto 1$.
\begin{figure}
    \begin{center}
    \includegraphics[angle=0, width=0.3\textwidth]{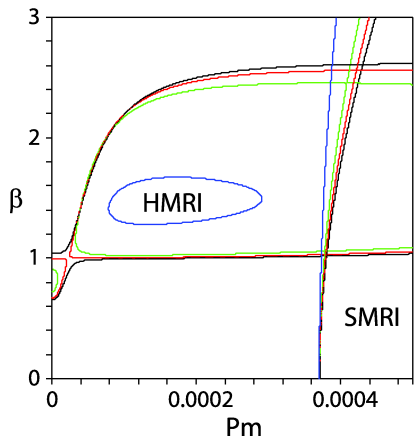}
    \end{center}
    \caption{SMRI and HMRI for $n=0$, ${\rm Ro}=-3/4$, ${\rm Ha}=30$.
       Black: ${\rm Rb} =-0.74$, ${\rm Re}\approx19876$, red: ${\rm Rb}=-0.75$,
       ${\rm Re}\approx21294$, green: ${\rm Rb}=-0.77$, ${\rm Re}\approx23935$, blue:
       ${\rm Rb}=-0.91$, ${\rm Re}\approx38553$, middle of the blue curve
       ${\rm     Pm}=0.0002$,
       $S={\rm Ha} {\rm Pm}^{1/2}\approx0.42$, ${\rm Rm}={\rm Re} {\rm Pm}\approx7.71$.}
    \label{fig2}
    \end{figure}
In Figure 3 we show that this mechanism is not restricted to $n=m=0$
but can easily extend to the range of AMRI with
higher azimuthal wavenumbers $m$, both for
small absolute values (Fig. 3a) and large absolute values (Fig. 3b) of $\beta$.
\begin{figure}
    \begin{center}
    \includegraphics[angle=0, width=0.45\textwidth]{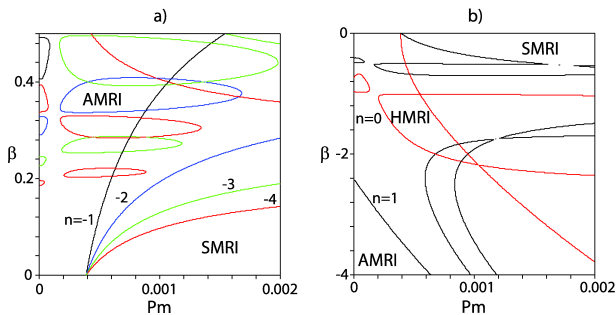}
    \end{center}
    \caption{Domains of SMRI, HMRI, and AMRI for ${\rm Ha}=30$, ${\rm Ro}=-3/4$, ${\rm Re}=4000$, ${\rm Rb}=-0.755$, and (a) black: $n=-1$, blue: -2, green: -3, red: -4 and (b) red: $n=0$, black: 1.}
    \label{fig2}
    \end{figure}

In summary, we have found that the range of applicability of the
inductionless versions of  MRI that were previously thought to be  restricted to
${\rm Ro}<{\rm Ro}_{\rm LLL}\approx-0.828$,
can easily extend to Keplerian profiles if only
${\rm Rm}$ is large enough to produce a $B_{\phi}(r)$ profile
that is somewhat \emph{shallower} than $1/r$.
Interestingly, the ${\rm Ro}_{\rm opt}^{+}({\rm Rb})$ curve starting with
the ULL further
continues to meet the ${\rm Ro}_{\rm opt}^{-}({\rm Rb})$ branch at the solid body rotation.
Since this
extension of the inductionless forms of MRI
circumvents the usual demand $S \propto 1$,
our finding may have significant consequences for the
working of MRI in the colder parts of accretion disks.
A detailed investigation of the respective roles of $S$ and $\rm Rm$ for the          
onset and the saturation mechanism of the instability                          
in different astrophysical problems goes beyond the scope of this letter and  must be   
be left for future work.                                            
We only note here that the sensitive structure of the instability
domains in the low $\rm Pm$-region, as seen in Figs. 2 and 3,
may easily trigger a quasi-oscillatory behaviour in the
non-linear regime.
Our results encourage experiments on the combination of MRI
and current driven instabilities as they are presently planned in the
framework of the DRESDYN project \cite{DRESDYN}.

This work was supported by Helmholtz-Gemeinschaft Deutscher
Forschungszentren (HGF) in frame of the Helmholtz Alliance LIMTECH,
as well as by Deutsche
Forschungsgemeinschaft in frame of the SPP 1488 (PlanetMag). We
acknowledge fruitful discussions with Marcus Gellert,
Rainer Hollerbach, and G\"unther R\"udiger.

\end{document}